\documentclass[preprint,showpacs,preprintnumbers,amsmath,amssymb,superscriptaddress,nofootinbib]{revtex4-2}
\usepackage{graphicx}
\usepackage{epstopdf}
\usepackage{dcolumn}
\usepackage{bm}
\usepackage{amssymb}
\usepackage{amsmath}
\usepackage{bm}
\usepackage{color}
\usepackage{lipsum}
\usepackage{times}
\usepackage{color}
\usepackage{xcolor}
\usepackage{hyperref}

\newcommand{\muB}{$\mu_\mathrm{B}$}

\newcommand{\bco}{BeCr$_2$O$_4$}
\newcommand{\bao}{BeAl$_2$O$_4$}
\newcommand{\cro}{Cr$_2$O$_3$}

\begin{document}

%
%
\title{Sawtooth lattice multiferroic \bco: Non-collinear magnetic structure and multiple magnetic transitions}
\author{Hector Cein Mandujano}
\affiliation{Department of Physics, 500 W University Ave, University of Texas at El Paso, TX 79968, USA}
\author{Alejandro Metta}
\affiliation{Department of Chemistry, 500 W University Ave, University of Texas at El Paso, TX 79968, USA}
\author {Neven Barišić}
\affiliation{Institute of Solid State Physics, Vienna University of Technology, Wiedner Hauptstr{\ss}e 8--10, 1040 Vienna, Austria}
\affiliation{Department of Physics, Faculty of Science, University of Zagreb, Bijenička cesta 32, 10000 Zagreb}
\author{Qiang Zhang}
\affiliation{Neutron Scattering Division, Oak Ridge National Laboratory, Oak Ridge, Tennessee 37830, USA.}
\author{Wojciech Tabiś}
\affiliation{AGH University of Science and Technology, Faculty of Physics and Applied Computer Science, 30-059 Krakow, Poland}
\author{Naveen Kumar Chogondahalli Muniraju}
\affiliation{The Henryk Niewodniczanski Institute of Nuclear Physics, Polish Academy of Sciences, 152 Radzikowskiego Str., 31-342, Krakow, Poland}
\affiliation{Institute of Solid State Physics, Vienna University of Technology, Wiedner Hauptstr{\ss}e 8--10, 1040 Vienna, Austria}
\affiliation{Institute of Physics, Bijeni\v{c}ka cesta 46, HR-10000, Zagreb, Croatia}
\author{Harikrishnan S. Nair}
\affiliation{Department of Physics, 500 W University Ave, University of Texas at El Paso, TX 79968, USA}
\date{\today}
\begin{abstract}
\noindent 
Noncollinear magnetic structures and multiple magnetic phase transitions in a sawtooth lattice antiferromagnet consisting of Cr$^{3+}$ are experimentally identified in this work, thereby proposing the scenario of magnetism-driven ferroelectricity in a sawtooth lattice. The title compound, \bco, displays three magnetic phase transitions at low temperatures, at $T_{N1}\approx$ 7.5~K, at $T_{N2}\approx$ 25~K and at $T_{N3}\approx$ 26~K, revealed through magnetic susceptibility, specific heat and neutron diffraction in this work. These magnetic phase transitions are found to be influenced by externally applied magnetic fields. Isothermal magnetization curves at low temperatures below the magnetic transitions indicate the antiferromagnetic nature of \bco\ with two spin-flop-like transitions occurring at $H_{c1}\approx$ 29~kOe and $H_{c2} \approx$ 47~kOe. Our high-resolution X-ray and neutron diffraction studies, performed on single crystal and powder samples unambiguously determined the crystal structure as orthorhombic $Pbnm$. By performing the magnetic superspace group analysis of the neutron diffraction data at low temperatures, the magnetic structure in the temperature range $T_{N3,N2}<T<T_{N1}$ is determined to be the polar magnetic space group, $P21nm.1^{\prime}(00g)0s0s$ with a cycloidal magnetic propagation vector $\textbf{k}_1$~=~(0, 0, 0.090(1)). The magnetic structure in the newly identified phase below $T_{N1}$, is determined as $P21/b.1^{\prime}[b](00g)00s$ with the magnetic propagation vector $\textbf{k}_2$~=~(0, 0, 0.908(1)). The cycloidal spin structure determined in our work is usually associated with electric polarization, thereby making \bco\ a promising multiferroic belonging to the sparsely populated family of sawtooth lattice antiferromagnets.
\end{abstract}
\maketitle

\section{Introduction}
\label{intro}
\indent
Single phase multiferroics attract fundamental and technological interest due to the fact that they combine switchable electric and magnetic polarization \cite{cheong2007multiferroics, spaldin2005renaissance, spaldin2019advances, khomskii2006multiferroics, wang2009multiferroicity}, and consequently offer the potential to cross-couple them and positively impact the field of spintronics \cite{bea2008spintronics}. 
Simply by cross-coupling of electric and magnetic degrees of freedom that multiferroics offer, one can efficiently control and thus manipulate magnetism by electric field or polarization current, which can provide a path to overcome bottlenecks in spintronics. 
One of the earliest experimental success along this line is the discovery of magnetoelectric (ME) effect in \cro\ in 1960s \cite{dzyaloshinskii1960magneto, astrov1960magnetoelectric}. 
Subsequently, a large pool of compounds have been identified and confirmed as multiferroics, including perovskites \cite{kimura2003magnetic}, delafossites \cite{terada2014spin}, charge-order systems \cite{van2008multiferroicity}, lone-pair systems \cite{wang2003epitaxial}, hexagonal manganites \cite{van2004origin}, pyroxenes \cite{jodlauk2007pyroxenes}, rare-earth chromates \cite{sahu2007rare}, molecule-based systems \cite{pardo2012multiferroics}, just to name a few. 
Exhaustive lists and details can be found in the review articles on multiferroics \cite{wang2009multiferroicity, spaldin2019advances, fiebig2016evolution, khomskii2006multiferroics}.
Multiferroics are classified into type I and II based on the strength of the coupling between magnetic and ferroelectric order parameters \cite{khomskii2009trend}. 
In type I multiferroics, the magnetic and ferroelectric ordering are well-separated in temperature and hence allows only for a weak coupling between the electric and the magnetic degrees of freedom. 
The type II members, on the other hand, generally develop ferroelectricity concomitantly with the magnetic order which is often established at low temperatures. 
Hence, the coupling between the two orders is more strong in the latter case.
Since magnetic ordering precedes the development of electric polarization in type II multiferroics, a lot of attention has been devoted to understanding how various types of spin order can lead to the development of electric polarization through the breaking of the inversion symmetry. 
Based on symmetry analyses it was pointed out that spiral magnetic order breaks the spatial inversion symmetry and generates electric polarization \cite{kenzelmann2005magnetic, katsura2005spin, sergienko2006role, mostovoy2006ferroelectricity}. 
Such approaches successfully explained the appearance of ferroelectricity in several families of spinels \cite{wang2019high}, perovskites \cite{kimura2003magnetic} and delafossites \cite{terada2014spin}.
Most of these compounds have three-dimensional magnetic lattices.
Low dimensional counterparts are less-explored, for example, there is only a single report of ferroelectricity induced by noncollinear spin order in sawtooth lattice compounds \cite{newnham1978magnetoferroelectricity}. 
\\
\indent 
A sawtooth lattice of spins is low-dimensional and frustrated. 
It is a model lattice to study complex quantum phases which was extensively explored in case of spin half systems \cite{kubo1993excited, sen1996quantum}. 
The sawtooth model is well described by the Hamiltonian, $\mathcal{H} = \sum_{i} \textbf{\emph{JS}}_i\cdot \textbf{\emph{S}}_{I+2} + \textbf{\emph{J}}'(\textbf{\emph{S}}_i\cdot \textbf{\emph{S}}_{i+1} + \textbf{\emph{S}}_{i+1}\cdot \textbf{\emph{S}}_{i+2}) - h\cdot \textbf{\emph{S}}_i$, where $\textbf{\emph{S}}_i$ usually represents a $S$~=~1/2 spin at site $i$ which has a neighbouring site $i~+~2$ in the spine of the sawtooth. $\textbf{\emph{J}}'$ is the interaction between the spine sites and the sawtooth tips and $h$ is the external magnetic field. 
Known as the $\Delta$ chain, delafossites, euchroite, metalorganic and fluorites are spin systems which have been studied in the context of sawtooth chains 
\cite{le2005first, kikuchi2011spin, inagaki2005ferro, baniodeh2018high, jeschke2019kagome}.
\\
\indent
After the pioneering work on chromate and olivine magnetoelectrics (i.e., Cr$_2$O$_3$ and \bco) in the '60s, it was only a decade ago that another olivine compound, Mn$_2$GeO$_4$, which was studied in the context of multiferroicity \cite{white2012coupling}. 
Similar to \bco, this compound also crystallizes in the orthorhombic $Pbnm$ space group and possesses a sawtooth lattice for the transition metal cation.
Interestingly, in Mn$_2$GeO$_4$ the spontaneous magnetization and electric polarization develop in the same crystallographic direction, below about 5.5~K \cite{white2012coupling, honda2012structure}. 
A multi-$\bf Q$ magnetic order with a commensurate (C) propagation vector $\textbf{k}_\mathrm{C}$~=~(0,~0,~0) and an \mbox{incommensurate} (IC) propagation vector $\textbf{k}_\mathrm{IC}$~=(0.136,~0.211,~0) are identified \cite{honda2017coupled}. 
The electric polarization in Mn$_2$GeO$_4$ is understood to originate from spiral spin order where ferromagnetic and ferroelectric domains synchronously switch. 
A recent work on another sawtooth magnetic lattice compound focuses on the atacamite compound Cu$_2$Cl(OH)$_3$ \cite{heinze2021magnetization}. 
Using high-magnetic field experiments, a complex evolution of magnetization was traced out in the atacamite and a plateau-like feature was observed at 315~kOe.
\\
\indent
The importance of sawtooth magnetic lattice extends beyond quantum magnetism of the $\Delta$ chain or the multiferroicity of Mn$_2$GeO$_4$. 
For example, the oxyselenite Fe$_2$O(SeO$_3$)$_2$ is reported to exhibit flat-band dynamics and phonon anomalies \cite{gnezdilov2019flat}. 
Flat-band systems with localized magnon states in high magnetic fields are  studied recently due to the possibility of dissipation-less magnonics \cite{bergholtz2013topological, leykam2018perspective, vicencio2015observation}. Strongly frustrated interactions within the sawtooth chains and weak interlayer coupling have been reported in oxy-arsenate Rb$_2$Fe$_2$O(AsO$_4$)$_2$ \cite{garlea2014complex}. 
Olivine-type chalcogenides, Fe$_2$Ge$Ch_4$ ($Ch$ = S, Se, Te) are computationally predicted as thermoelectrics \cite{gudelli2015predicted} and have been observed to reveal frustration-driven complex low-temperature magnetic phases \cite{nhalil2019antiferromagnetism}.\\
\indent
The magnetic susceptibility of \bco\ was first reported by Santoro~{\em et al.} \cite{santoro1964magnetic}. 
According to their report, \bco\ had a N{\'e}el temperature of 28~K, Curie-Weiss temperature of 13~K, and an effective magnetic moment of 3.2~\muB\  per Cr. 
A four-sublattice Weiss model was proposed to explain the magnetism of \bco. 
However, antiferromagnetic resonances at 29.5~kG and 47.7~kG suggestive of non-collinear spin arrangement \cite{elliston1967antiferromagnetic}, and spin-flops were later reported \cite{ranicar1967spin}. 
The spiral magnetic structure was then verified by neutron diffraction study which estimated the periodicity as 65~{\AA} \cite{cox1969neutron}. 
The spiral arrangement of spins breaks inversion symmetry and leads to the generation of ferroelectricity, which was experimentally observed as a weak effect, without corresponding anomalies in dielectric permittivity or electrical conductivity at the N{\'e}el temperature \cite{newnham1978magnetoferroelectricity}.
In this work we present detailed magnetic and thermodynamic characterization of \bco, mapping out a double-transition in magnetic susceptibility and specific heat at $T_{N3}\approx$ 26~K, at $T_{N2}\approx$ 25~K, in addition to a further low-temperature anomaly at $T_{N1}\approx$ 7.5~K. 
To the best of our knowledge this is the first time, a cascade of three magnetic transitions are reported in \bco. 
Based on the results from our measurements, we have compiled the $H$-$T$ phase diagram of \bco\ in magnetic fields up to 80~kOe. 
We have determined the magnetic structure of \bco\ through neutron powder diffraction experiments and subsequent analysis using the magnetic superspace group approach.
The magnetic structure at 2~K is an incommensurate structure with a propagation vector (0,~0,~0.908(1)), in a monoclinic supercell setting.
The results presented here should motivate a re-examination of ferroelectric polarization, and magnetization under high magnetic fields complemented with magnetic structure determination of \bco, especially using single crystals.

\section{Methods}
\subsection{Experimental techniques}
\label{expt}
The powder samples of \bco\ used in the present work were prepared by solid state synthesis method by reacting BeO and Cr$_2$O$_3$ (4N purity, Sigma Aldrich) in stoichiometric ratios at 1400~$^\circ$C. 
The precursor oxides were mixed and ground using a mortar and pestle and heated up to five times, with intermediate grinding after each heating cycle that lasted for 24~h. 
The resulting powder samples were dark green in color and were characterized for quality by taking laboratory-based powder diffraction patterns using Cu K$\alpha$ X-rays of wavelength 1.54 {\AA} (Panalytical Empyrean). 
A crystal growth experiment was attempted using the method of optical floating zone.
Although this did not result in the congruent melting of \bco, thin whisker-shaped crystals were obtained from this attempt.
Using those crystals, room temperature single crystal X-ray diffraction data was collected by a Bruker Quest Photon 200 diffractometer using the Apex3 suite. 
The crystal structure was solved and refined using the Bruker SHELXTL software package \cite{sheldrick1997shelxtl}. 
High-resolution synchrotron X-ray diffraction experiments were performed on powder samples of \bco\ at beamline 11-BM,  Advanced Photon Source, Argonne National Lab. 
Synchrotron generated X-rays of wavelength 0.4811~{\AA} were used to collect diffraction patterns at 90~K and 295~K. 
Magnetic properties of \bco\ pellets were established using a Magnetic Property Measurement System (MPMS) SQUID magnetometer. 
Magnetic susceptibility and isothermal magnetization were measured in the temperature range 1.8~K to 320~K in magnetic fields up to 90~kOe. 
The temperature dependence of specific heat was measured in zero and applied magnetic fields up to 40~kOe in a Physical Property Measurement System (PPMS). 
Neutron powder diffraction patterns were collected at the high-resolution time-of-flight diffractometer, POWGEN \cite{powgen} at Spallation Neutron Source, Oak Ridge National Laboratory, at selected temperatures in the range 1.8~K to 200~K using two data banks with center wavelengths of 1.5~{\AA} and 2.665~{\AA}. 
About 4~grams of \bco\ powder was loaded in a 6~mm diameter vanadium canisters for those measurements.
Analysis of X-ray and neutron diffraction data from 11-BM and POWGEN was performed using FullProf suite of programs \cite{fullprof} and JANA2020 \cite{jana}. The graphical representations of lattice and magnetic structures used in this work are created using the program VESTA \cite{momma2011vesta}.

\section{Results and discussion}
\subsection{Crystal structure}
\label{sec-str}
\begin{figure*}[!t]
	\centering
	\includegraphics[scale=0.5]{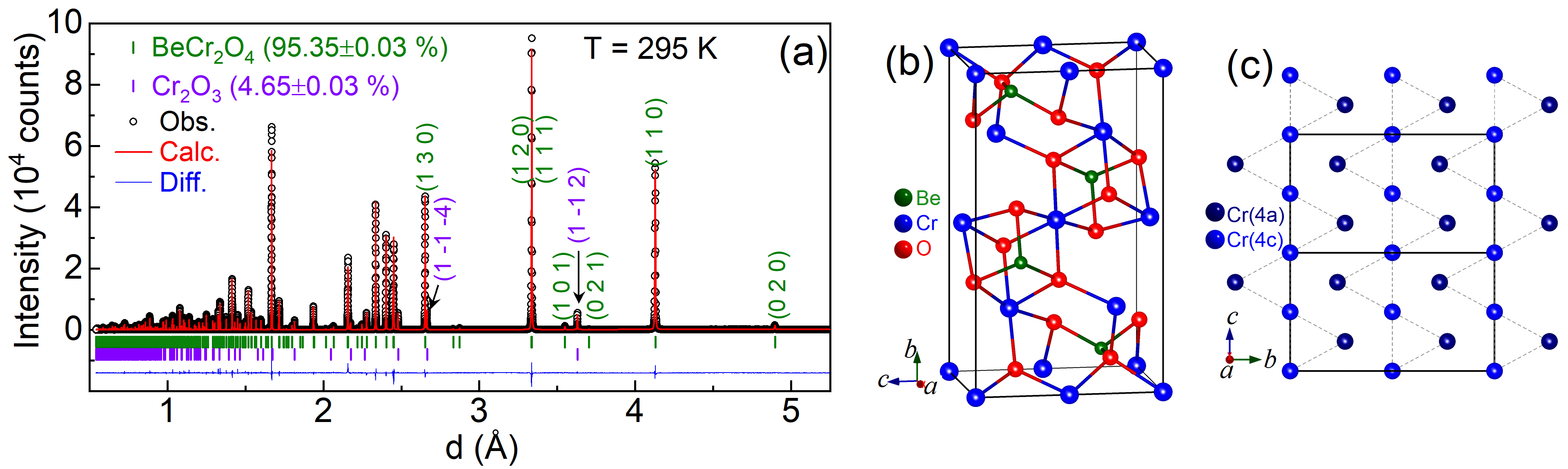}
	\caption{(a) Synchrotron X-ray diffraction pattern of BeCr$_2$O$_4$ at 295~K along with Rietveld fit using $Pbnm$ space group. Circles are the experimentally measured intensities and the solid line is the calculated pattern. Top and bottom vertical bars mark the positions of the expected Bragg reflections for BeCr$_2$O$_4$ and Cr$_2$O$_3$, respectively. Blue horizontal line at the bottom is the difference between the measured and calculated patterns. Indices of few major reflections are indicated. (b) A clinographic view of the orthorhombic unit cell of BeCr$_2$O$_4$. (c) The sawtooth arrangement formed by Cr atoms at $4a$ and $4c$ Wyckoff positions. The solid line indicates the unit cell boundary.}
	\label{fig-11bm}
\end{figure*}

The phase-purity and the crystal structure of the powder samples of \bco\ were checked first using the X-ray powder diffraction patterns obtained using a laboratory diffractometer. 
It was confirmed to crystallize in the orthorhombic space group, $Pbnm$, showing no impurity peaks.
The crystal structure was further studied using high-resolution synchrotron X-ray diffraction data acquired from 11-BM instrument at APS. 
The diffraction pattern at 295~K along with the Rietveld refinement is shown in Figure~\ref{fig-11bm}(a). 
We confirmed the preliminary assignment of the orthorhombic $Pbnm$ space group \cite{cox1969neutron,yamnova2014specific} for \bco\ through the refinement of the high-resolution synchrotron X-ray diffraction data. \\
\begin{table}[!htb]
	\renewcommand{\tabcolsep}{2.4mm}
	\caption{
		The refined fractional atomic positions of \bco\ obtained from the Rietveld refinement of synchrotron data (11-BM) shown in Figure~\ref{fig-11bm}. The orthorhombic space group with $Pbnm$ setting was used for the refinements, the refined lattice parameters at $T$ = 295~K are
		$a$ = 4.5550(1)~{\AA},
		$b$ = 9.7924(2)~{\AA} and
		$c$ = 5.6651(1)~{\AA}.
		The goodness-of-fit ($\chi^2$) was 1.85, R$_\mathrm{wp}$ = 9.88$\%$ and {R$_\mathrm{exp}$} = 5.34$\%$. The bond distances and the angles are also shown in this table. BVS stands for bond valence sums, here estimated for Be and Cr.}
	\label{tab-11bm}
		\begin{tabular}{llllll} \hline \hline
			& \textit{x} & \textit{y} &\textit{z} & S.O.F & B$_\mathrm{iso}$ ({\AA}$^{2}$)\\ \hline
			Be ($4c$) & 0.5681(1) & -0.0929(6)  & 0.75  & 1.00      & 0.65(4) \\
			Cr ($4a$) & 0.0       &  0.0        & 0.0   & 0.933(6)  & 0.13(1) \\
			Cr ($4c$) & 0.4974(3) & 0.2311(6)   & 0.75  & 0.931(6)  & 0.11(4) \\
			O1 ($4c$) & 0.2768(6) & 0.4072(3)   & 0.75  & 0.994(9)  & 0.40(4) \\
			O2 ($4c$) & 0.7289(6) & 0.0612(3)   & 0.75  & 1.000(6)  & 0.45(3)\\
			O3 ($8d$) & 0.2587(3) & 0.1637(3)   & 0.0221(3) & 0.989(9) & 0.48(4) \\ \hline
		\end{tabular}
		\begin{tabular}{lll|lll}
			\multicolumn{3}{c}{Bond distance ({\AA})} & \multicolumn{3}{c}{Bond angle ($^\circ$)}  \\ \hline
			\multicolumn{2}{c}{Cr($4a$)-Cr($4c$)} & 3.5021(4) & \multicolumn{2}{c}{O($4c$)-Cr($4a$)-O($8d$)} & 83.64(3) \\
			\multicolumn{2}{c}{Cr($4a$)-Cr($4a$)} & 5.3995(1) & \multicolumn{2}{c}{O($4c$)-Cr($4a$)-O($8d$)} & 96.37(4) \\
			\multicolumn{2}{c}{Cr($4a$)-Cr($4a$)} & 9.7923(2) & \multicolumn{2}{c}{O($4c$)-Cr($4a$)-O($4c$)} & 93.06(4) \\
			\multicolumn{2}{c}{Cr($4c$)-Cr($4c$)} & 3.6533(2) & \multicolumn{2}{c}{Cr($4c$)-O($8d$)-Cr($4a$)} & 95.78(4) \\
			\multicolumn{2}{c}{Cr($4c$)-Cr($4c$)} & 5.3901(2) & \multicolumn{2}{c}{Cr($4c$)-O($4c$)-Cr($4a$)} & 126.29(4) \\
			\multicolumn{2}{c}{Be($4c$)-O($8d$)} & 1.664(2) & \multicolumn{2}{c}{Cr($4a$)-O($4c$)-Cr($4a$)} & 92.20(5) \\
			\multicolumn{2}{c}{Be($4c$)-O($4c$)} & 1.571(3) & \multicolumn{2}{c}{Cr($4c$)-O($8d$)-Cr($4c$)} & 129.82(5) \\ \hline
		\end{tabular}
		\begin{tabular}{llllll}
			&\multicolumn{1}{c}{BVS} & & Be (1.981(4)) & Cr (3.296(2)) & Cr (3.030(2)) \\ \hline
		\end{tabular}
\end{table}
\indent 
\bco\ adopts an olivine-based crystal structure similar to that of the mineral Mg$_2$SiO$_4$. 
In this structure, Be$^{2+}$ occupies a tetrahedral position and Cr$^{3+}$, octahedral. 
An early study involving neutron diffraction had estimated the lattice parameters, $a$ = 4.555~{\AA}, $b$ = 9.792~{\AA} and $c$ = 5.663~{\AA} for \bco\ in $Pbnm$ space group \cite{cox1969neutron}. 
Mineral \bco\ found in Russia was reported in orthorhombic $P2_12_12_1$ space group which is a subgroup of acentric $Pcmn$ \cite{yamnova2014specific}, and also in $Pnma$ \cite{pautov2013}. 
The structural parameters refined using the synchrotron X-ray diffraction data (Figure~\ref{fig-11bm}) in the present work are given in Table~\ref{tab-11bm}. 
The Cr atoms at the $4a$ and $4c$ Wyckoff positions form a sawtooth lattice which is shown schematically in Figure~\ref{fig-11bm}(c). 
This sawtooth lattice is constituted by the Cr atoms which form isosceles triangles with two different Cr--Cr bond distances, 2.9936(6)~{\AA} and 2.8367(4)~{\AA}. 
The bond distances and angles are listed in Table~\ref{tab-11bm}. 
The edge-sharing Cr atoms are found to be about 2.83~{\AA} apart \cite{santoro1964magnetic}. 
The Cr $4a$ position has local inversion symmetry while the $4c$ position has a mirror symmetry. 
The four-sublattice model proposed for \bco \cite{santoro1964magnetic} takes into account the direct and superexchange pathways between Cr atoms at the inversion and mirror symmetry points. 
The Cr--O--Cr bond angle of 129$^\circ$ is more favourable to superexchange. 
The bond valence sums \cite{bondvalence} determined from the analysis of the present crystal structure data indicate that the valence states of the cations in \bco\ are Be$^{1.98}$, Cr$^{3.29}$ and Cr$^{3.03}$.
We did not observe any incommensurate lattice modulations as reported by another group \cite{kovalev1995effect}. 
The unit cell obtained from the present study is similar to the mineral mariinskite \cite{yamnova2014specific,weir1960studies} in which the Cr atoms have two independent positions and Be has only one position, both of which are in a distorted octahedral environment surrounded by oxygen atoms.
The crystal structure solution using the single crystal X-ray diffraction data confirmed orthorhombic $Pbnm$ space group that we arrived at through powder diffraction data analysis. 
The results of crystal diffraction experiments are presented in Table ~\ref{tab-sxrd}.
\begin{table}
	\renewcommand{\tabcolsep}{2.4mm}
	\caption{The sample and crystal data as well as the details of data collection and refinements from the single crystal structure analysis of \bco. } \vspace{0pt}
	\label{tab-sxrd}
	\begin{tabular}{llll} \hline \hline
		Chemical formula & 	\bco\ \\
		Formula weight~(g.mol$^{-1}$)	& 177.01  \\
		Temperature~(K)	& 296(2) \\
		Wavelength~({\AA}) & 0.71073 \\
		Crystal system	&  Orthorhombic \\
		Space group	& $Pbnm$ \\
		Unit cell dimensions & a~({\AA}) = 4.5576(2),	$\alpha$~($^\circ$) = 90 \\
		& b~({\AA})  = 9.7970(5),	$\beta$~($^\circ$)  = 90 \\
		& c~({\AA})  = 5.6658(3), 	$\gamma$~($^\circ$)  = 90 \\
		Volume~({\AA$^3$})   & 252.98(2)  \\	
		Z & 4 \\
		Calculated density (g.cm$^{-3}$) & 4.647  \\
		Absorption coefficient~(mm$^{-1}$) & 8.370  \\
		F(000) & 336 \\
		Theta range for data collection ($^\circ$)	& 4.16 to 60.01 \\
		Reflections collected & 13444 \\
		Independent reflections	& 2009 [R(int) = 0.0514] \\
		Coverage of independent reflections	& 99.4~\% \\
		Absorption correction & Multi-Scan \\
		Structure solution program	& XT, VERSION 2018/2 \\
		Refinement program	& SHELXL-2019/1 (Sheldrick, 2019) \\
		Function minimized	& $\sum$ w(F$^2_0$ - F$^2_c$)$^2$ \\
		Data / restraints / parameters	& 2009 / 0 / 41 \\
		Goodness-of-fit on F2 & 1.295 \\
		$\Delta$/$\sigma_\mathrm{max}$ & 0.001 \\
		Final R indices	& 1758 data; I$>$2$\sigma$(I),	R1 = 0.0404, wR2 = 0.1041 \\ 
		& all data, R1 = 0.0436, wR2 = 0.1084 \\ \hline
	\end{tabular}
\end{table}
\subsection{Magnetic properties: Multiple phase transitions}
\label{sec-magnetic}
\begin{figure*}[!t]
	\centering
	\includegraphics[scale=0.45]{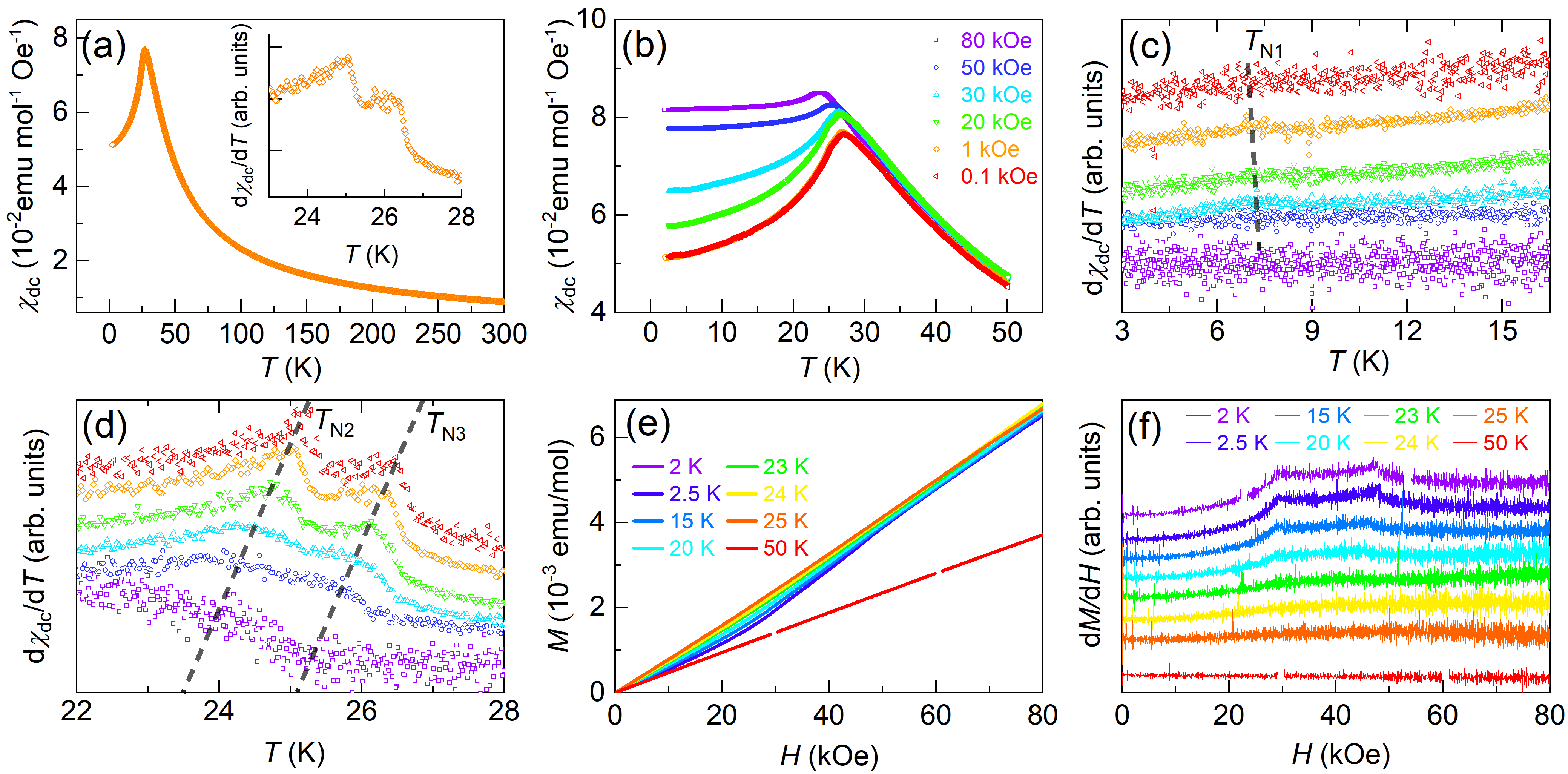}
	\caption{\label{fig-mag}
		(a) Magnetic susceptibility, $\chi_\mathrm{dc}(T)$, of \bco\ measured at an applied field of 1~kOe. The inset magnifies the derivative, d$\chi_\mathrm{dc}(T)/$d$T$ in the low temperature region, revealing a double-peak anomaly of two near-by transitions. (b) The magnetic susceptibility at different values of applied magnetic fields from 0.1~kOe to 80~kOe. (c, d) Derivatives d$\chi_\mathrm{dc}(T)/$d$T$ identifying three transitions at $T_{N1}\approx$ 7.5~K, $T_{N2}\approx$ 25~K and $T_{N3}\approx$ 26~K. (e) The isothermal magnetization $M(H)$ at different temperatures in the range 2~K to 50~K. (f) The derivative, d$M/$d$H$, revealing the presence of field-induced magnetic transitions.}
\end{figure*}

The bulk magnetic properties of \bco\ are shown in Figure~\ref{fig-mag}. 
The dc magnetic susceptibility ($\chi_\mathrm{dc}(T)$) was measured in the temperature range of 2 -- 300~K under the application of different magnetic fields. 
A plot of the magnetic susceptibility as a function of temperature measured at 1~kOe, is shown in Figure~\ref{fig-mag}(a). 
At high temperatures a paramagnetic response is seen in $\chi_\mathrm{dc}(T)$ which increases towards a prominent peak at around 26~K. 
The magnetic susceptibility further decreases beyond the peak in the low temperature region. 
The inset of (a) shows the plot of the derivative $d\chi_\mathrm{dc}(T)/dT$ versus temperature. 
Two prominent discontinuities at 25~K and 26~K are recovered from the derivative of magnetic susceptibility, signifying the presence of two near-by magnetic transitions. 
These features were not observed in previous studies reported on \bco.
The variation of low temperature ($T <$ 50~K) magnetic susceptibility with the application of different magnetic fields is shown in Figure~\ref{fig-mag} (b). 
The magnitude of the magnetic susceptibility is seen to increase with the application of higher values of magnetic fields. 
At 80~kOe, a near-saturation of low temperature susceptibility is observed.
%
%
Figure~\ref{fig-mag}(c,d) show the derivative $d\chi_\mathrm{dc}/dT$ in two different temperature ranges, 4 -- 16~K and 22 -- 28~K respectively, at different magnetic fields. 
A subtle low-temperature anomaly, designated as $T_{N1}$, has been identified at approximately 7~K. This anomaly is not highly pronounced (panel Fig~\ref{fig-mag}(c)), but it does coincide with the more noticeable anomaly observed in specific heat at a similar temperature. 
The double-peak anomalies at higher temperatures are designated $T_{N2}$ (25~K) and $T_{N3}$ (26~K). 
One of the earliest works on the magnetic susceptibility of \bco\ reported a broad feature at approximately 28~K \cite{cox1969neutron}. 
Although the qualitative features of the magnetic phase transition reported by Cox {\em et al.} are similar to the present work, our results clearly reveal three anomalies in the magnetic susceptibility of \bco\ for the first time.
\\
\indent
The magnetic susceptibility of \bco\ presented in Figure~\ref{fig-mag}(a) was analyzed using Curie-Weiss law by performing a least-squares fit to the data. 
An effective paramagnetic moment of 3.48(7)~\muB/Cr and a Curie-Weiss temperature, $\theta_\mathrm{CW}$ = --40~K are obtained from the fit. 
The paramagnetic moment value which we have obtained is comparable to, albeit slightly reduced, the spin-only value of Cr$^{3+}$, 3.88~\muB\ ($d^3$, $S$ = 3/2, $g\approx$2). 
In comparison, the reported effective moment is 3.2~\muB\ and $\theta_\mathrm{CW}$ = --13~K \cite{cox1969neutron, santoro1964magnetic}. 
Previous study of the bulk magnetic susceptibility of \bco\ dates back to 1963 \cite{santoro1964magnetic}. 
Since a microscopic picture of the magnetic structure of Cr$^{3+}$ was lacking at that time, a macroscopic model based on exchange interactions between the nearest-neighbour Cr atoms was put forward. 
As mentioned in the previous section on the crystal structure of \bco, two types of Cr atoms based on the local symmetry was identified as $m$ and $i$ and subsequently, two interaction strengths, $\alpha$ and $\beta$, corresponding to Cr$_m$-O-Cr$_i$ and Cr$_m$-O-Cr$_m$ respectively, were introduced. 
Using these parameters, a four-sublattice model was adopted to analyze the magnetic susceptibility. 
It was deduced that direct magnetic exchange between Cr$^{3+}$ moments is important in this compound. 
The magnetization isotherms obtained at different temperatures between 2~K to 50~K are shown in Figure~\ref{fig-mag}(e) and their derivatives $dM/dH$ in Figure~\ref{fig-mag}(f). 
In general, an antiferromagnetic-like response is seen at all temperatures, except a clear change in slope is observed below 25~K down to 2~K. 
The derivative-plot in Figure~\ref{fig-mag}(f) shows the presence of two anomalies at $H_{c1} \approx$ 29~kOe and $H_{c2} \approx$ 47~kOe which can be identified as the critical fields of spin-flop type field-induced transitions. 
The temperature-dependence of the critical fields $H_{c1}$ and $H_{c2}$ are shown in Figure~\ref{fig-cp} (c). 
\\
\indent
The atacamite Cu$_2$Cl(OH)$_3$ containing Cu sawtooth is a compound similar to \bco. 
It enters an ordered magnetic state at 8.5~K, as seen in the magnetic susceptibility at low fields \cite{heinze2021magnetization}.
The magnetic susceptibility measured at different external magnetic fields is seen to change very little even at 13~T with
$H \parallel b$-axis, which is identified as the easy axis.
The $S$ = 1/2 Cu$^{2+}$ magnetic lattice in the atacamite results in spin-flop transitions in two different values of magnetic fields, at approximately 4~T and at 30~T when the field is applied parallel to the $b$ axis.
These effects are reflected in magnetostriction also.
Although very interesting magnetization plateaus are present in the magnetization of the atacamite which is reminiscent of
quantum effects, the current explanation for this is offered in terms of strongly coupled 1D ferromagnetic clusters that are coordinated
by weak, anisotropic 2D interactions \cite{heinze2021magnetization}.
Another sawtooth magnetic lattice reported recently is that of Fe$_2$Se$_2$O$_7$ \cite{nawa2021magnetic}.
Different from the Cu sawtooth of the atacamite, the Fe sawtooth in this case orders magnetically
at high temperatures, the transition temperature being at 112~K.
The easy axis, in this case too, is the $b$-axis of an orthorhombic unit cell.
Spin-flop magnetic transitions are found to occur at 5.45~T when the field is applied parallel to the $b$-axis.
The spin chain system Rb$_2$Mn$_3$(MoO$_4$)$_3$(OH)$_2$, similar
to \bco, shows two nearby magnetic anomalies
in magnetic susceptibility
and specific heat \cite{liu2020complex}.
Magnetic susceptibility of this compound
revealed anomalies at 4.7~K and 3.2~K, while
a further low-temperature kink was seen in
specific heat at 2.6~K indicating a possible
change of magnetic structure.
A significantly high degree of frustration
was seen in the compound with $f$ = 24.
Fe$_2$O(SeO$_3$)$_2$ is another sawtooth lattice system with a relatively high temperature
transition at 120~K \cite{gnezdilov2019flat}.
A single-phase transition peak is observed
with low-temperature susceptibility nearly
saturating, indicating antiferromagnetic
correlations being present.
Spin-flop transition, common to the sawtooth systems, is indicated to be present at
low applied fields and, similar to the
atacamite, ferromagnetic correlations
are also speculated to be present.
Hence we see a gradual trend, with some exceptions, of increasing
phase transition temperature with increasing
value of spin as we look at sawtooth systems
made up of Cu, Cr, Mn and Fe spins.
Common to all these systems is the presence
of spin-flop magnetization transitions
and plateaus at high fields.
%
\subsection{Specific heat and the H-T phase diagram}
\label{sec-cp}
\begin{figure*}[!t]
	\centering
	\includegraphics[scale=0.45]{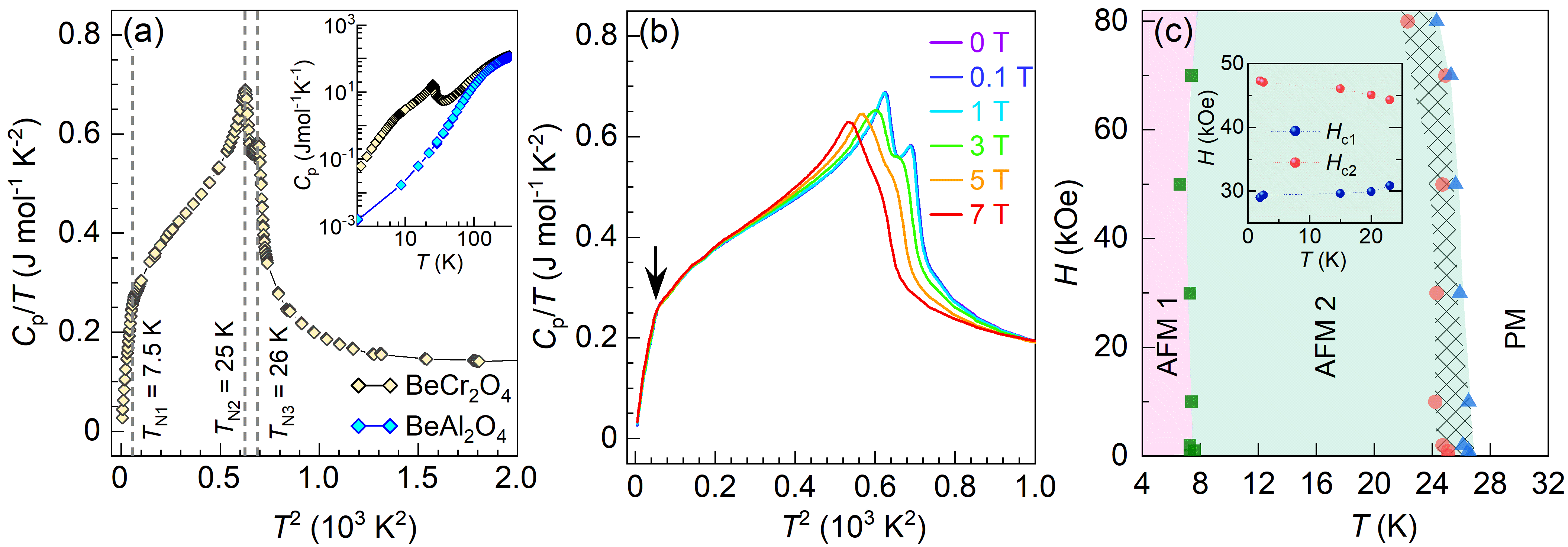}
	\caption{\label{fig-cp}(a) The low temperature specific heat of \bco\ plotted as $C_p/T$ versus $T^2$. Three phase transitions at $T_{N1}\approx$ 7.5~K, $T_{N2}\approx$ 25~K and $T_{N3}\approx$ 26~K are present which are indicated by dashed vertical lines. The inset shows the total specific heat of \bco\ and \bao\ in the range 2-300~K, in logarithmic scales. (b) The specific heat of \bco\ under the application of different magnetic fields. $T_{N2}$ and $T_{N3}$ are suppressed by the fields while $T_{N1}$ (indicated by an arrow) is field independent. (c) The $H$-$T$ phase diagram of \bco\ constructed from the phase transition points extracted from the specific heat and magnetic susceptibility. The inset shows the critical fields $H_{c1}$ and $H_{c2}$ identified in the derivatives of magnetization isotherms.}
\end{figure*}

The specific heat $C_p(T)$ of \bco\ presented in Figure~\ref{fig-cp}(a) shows three prominent phase transitions at 7.5~K, 25~K, and 26~K  thereby confirming the magnetic phase transitions at $T_{N1}$, $T_{N2}$ and $T_{N3}$ previously seen in the magnetic susceptibility of the compound. 
The plot of $C_p(T)$ versus $T^2$ shown in Figure~\ref{fig-cp}(a) highlights the anomalies at $T_{N1}$, $T_{N2}$ and $T_{N3}$ which are more pronounced peaks than in the magnetic susceptibility curves, Figure~\ref{fig-mag}(a).
%
%
The curves of total specific heat of \bco\ and that of the non-magnetic analogous compound \bao, measured in the range 2~K -- 320~K are shown in the inset of Figure~\ref{fig-cp}(a), plotted as $C_p(T)$ versus $T$, in logarithmic axes. 
The $C_p(T)$ of \bao\ was used as the phonon analog to be subtracted from that of \bco\ in order to obtain the magnetic entropy of Cr$^{3+}$ moments as 9~J~mol$^{-1}$~K$^{-1}$ at 27~K. 
Assuming $S$~=~3/2 for Cr$^{3+}$, the spin-only contribution to magnetic entropy can be estimated using the expression 2$\mathrm{R}$~ln(2$S$ + 1) = 23~J~mol$^{-1}$~K$^{-1}$. 
A curve fit according to the Debye model to the full temperature range of total specific heat of \bao\ resulted in the estimation of the  Debye temperature $\theta_D$ (\bao) = 992.01(1) K. 
The reported Debye temperatures of \bao\ and BeO are 1096~K and 1280~K respectively \cite{wang2019high, morell1996anharmonic}. 
The Debye temperature we obtained is close to the value obtained by analyzing the low-temperature specific heat of \bao\ using the expression, $C_\mathrm{low}$ = $\alpha T$ + $\beta T^3$ + $\gamma T^5$ where $\alpha$, $\beta$ and $\gamma$ are fitting parameters. 
The temperature range 2~K--100~K was used for the low-temperature fit, and a value of $\beta$ = 1.63~$\times$ 10$^{-5}$~Jmol$^{-1}$K$^{-4}$ was obtained. 
The Debye temperature was estimated using the expression, $\theta_\mathrm D$ = $\left( \frac{12p\pi^4 R}{5\beta}\right)$ = 941.13(2)~K, where $p$ is the number of atoms in the formula unit and $R$ is the universal gas constant. 
Due to the presence of multiple low-temperature anomalies in \bco, an analysis of specific heat using the $T^3$-law was not feasible. 
We applied the molecular mass correction for the Debye temperature following [MV$^{2/3}$$\theta^2_\mathrm D$](\bao) = [MV$^{2/3}$$\theta^2_\mathrm D$](\bco) where $V$ is the unit cell volume, to obtain $\theta_\mathrm D$ (\bco)~=~770~K. 
The total specific heat of \bco\ was fitted with the Debye model \cite{kittel1996introduction} for phonon-specific heat which leads to a Debye temperature, $\theta_\mathrm D$ (\bco) = 847~K. 
\\
\indent 
The phase transitions at $T_{N1}$, $T_{N2}$ undergo appreciable changes with the application of external magnetic fields to a value up to 70~kOe. It can be seen from Figure~\ref{fig-cp} (b) that a magnetic field of 50~kOe or 70~kOe smoothed the twin-peak near $T_{N2}$ and $T_{N3}$,
but are shifted towards low temperatures very slightly. 
The signature of $T_{N1}$ is indicated by the arrow in the figure
where a change of slope is visible.
A magnetic field-temperature (HT) phase diagram of \bco\ presented as Figure~\ref{fig-cp}(c), was constructed from the phase transition points obtained from the derivatives of magnetic susceptibility and specific heat curve.
The phase diagram and the magnetic properties deduced from the bulk physical properties measurements can be summarized as follows.
At temperatures above $\approx$ 30~K, \bco\ is paramagnetic. 
When the temperature is reduced below 30~K three magnetic transitions are observed with peaks at temperatures $T_{N3}$, $T_{N2}$ and $T_{N1}$ ($T_{N3} >T_{N2} > T_{N1}$). 
The double-peak high-temperature transitions ($T_{N3}$ and $T_{N2}$) occur within a Kelvin apart, the magnetic state below these transitions is expected to be antiferromagnetic, designated in our HT phase diagram as AFM~2. 
Further, below $T_{N1}$, we propose another antiferromagnetic phase AFM~1. 
The microscopic nature of the magnetic order in the narrow temperature region between $T_{N3}$ and $T_{N2}$ and below $T_{N1}$ is yet to be determined. 
In the subsequent sections, we will show results from neutron diffraction experiments to probe the magnetic structures of AFM~1 and AFM~2.
%
\subsection{Neutron diffraction: Noncollinear magnetic structure}
\label{sec-npd}
\begin{figure*}[!t]
	\centering
	\centering\includegraphics[scale=0.47]{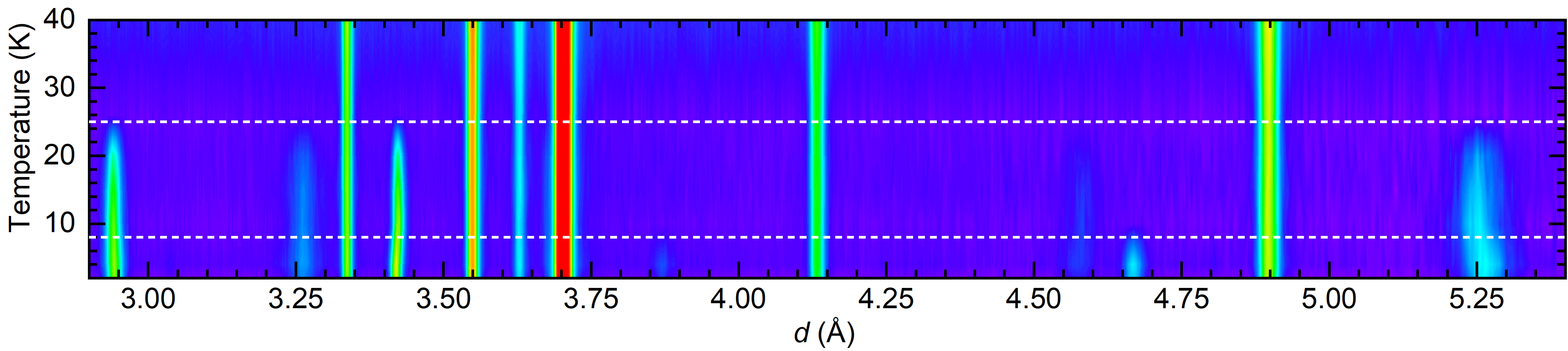}
	\caption{\label{fig-npd1} (color online) A color map of temperature evolution of diffraction patterns of \bco\ as a function of $d$-spacing. Magnetic phase transitions, at approximately $T_{N2,N3}$ (top horizontal dashed line) and $T_{N1}$ (bottom horizontal line), are discernible with the appearance of additional Bragg reflections.}
\end{figure*}
Based on the results of our magnetization and specific heat measurements presented in the previous sections, three magnetic phase transitions are identified in \bco\ at $T_{N1}$, $T_{N2}$ and $T_{N3}$. 
To elucidate the microscopic lattice and magnetic ordering at different temperatures we performed neutron powder diffraction experiments in the range 2 -- 295~K. 
The combined diffraction data obtained at several temperatures is presented in figure~\ref{fig-npd1} in the form of a color map of the diffracted intensity, plotted as a function of temperature and $d$-spacing in the range 2.9~--~5.4~{\AA}. 
Distinct transitions are discernible with the appearance of new Bragg reflections at $\sim$ 26~K and $\sim$ 7~K. 
Two horizontal dashed lines are marked in figure~\ref{fig-npd1} to indicate the temperatures below which additional Bragg reflections are observed. 
The top line (marked as $T_{N2,N3}$) is close to the transitions $T_{N3}$ and $T_{N2}$, while the bottom line corresponds to the transition at $T_{N1}$. 
Due to the lack of finer temperature steps in the current neutron data, the magnetic ordering in the temperature range $T_{N3} > T > T_{N2}$ (the hashed region in Figure~\ref{fig-cp} (c)) could not be resolved from our neutron powder diffraction measurements, so the phase transition around this temperature region is indicated as $T_{N2,N3}$ in the following section. \\
\indent
The neutron diffraction pattern collected at $T$ = 295~K was refined using the orthorhombic structure model (space group $Pbnm$) obtained from our X-ray diffraction data analysis. 
The lattice parameters obtained from the refinements are $a$ = 9.840(1)~{\AA}, $b$ = 5.693(7)~{\AA} and $c$ = 4.577(4)~{\AA}. 
These values are slightly different from the unit cell parameters determined using synchrotron X-ray diffraction experiments presented earlier but are in general agreement with the earlier reports~\cite{cox1969neutron}. 
The refined fractional coordinates at 295~K are comparable to the values obtained from the refinement of the synchrotron X-ray data given in Table~\ref{tab-11bm}.
\\
\begin{figure*}[!t]
	\centering
	\includegraphics[scale=0.47]{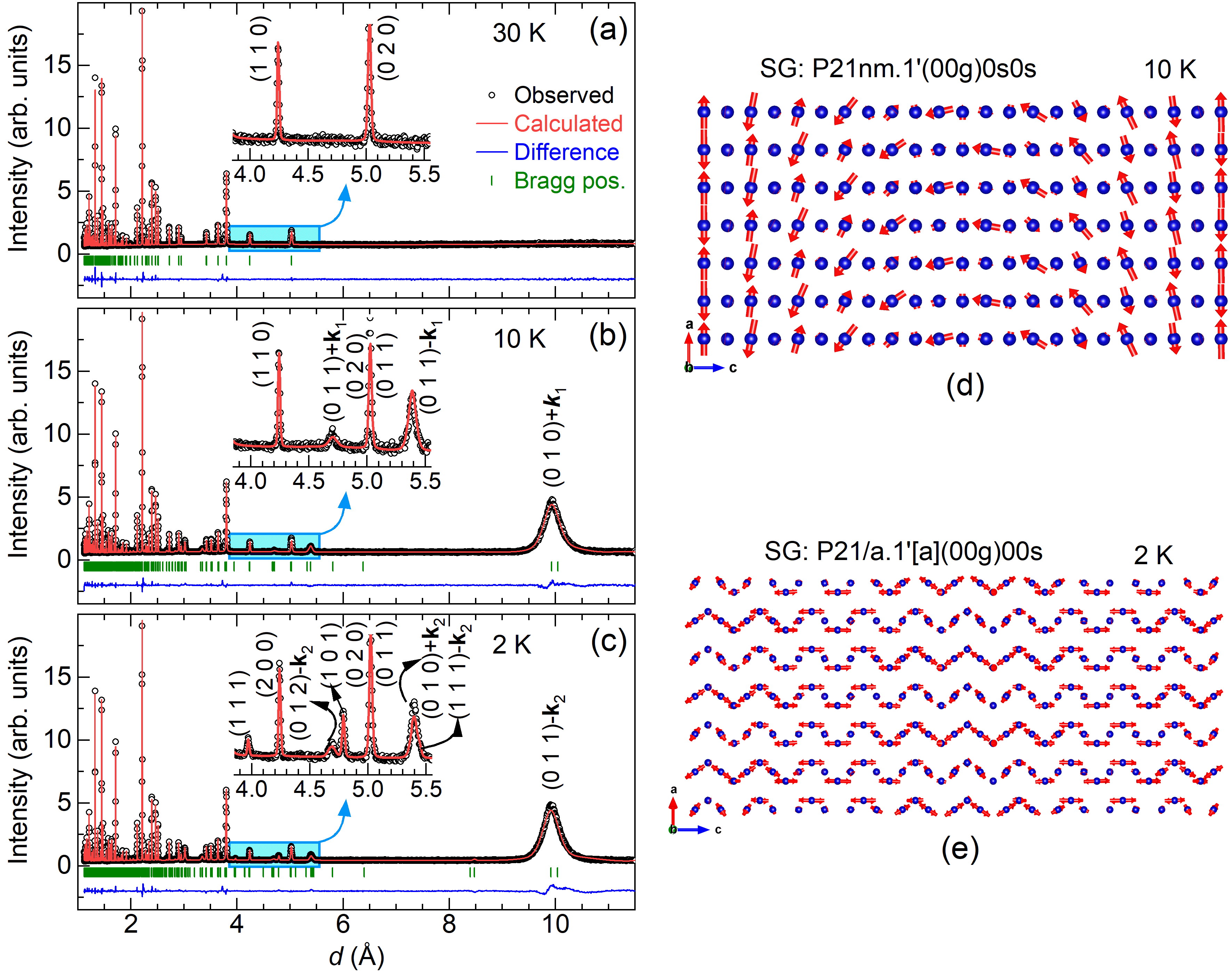}
	\caption{\label{fig-npd2} Time-of-flight neutron diffraction data of \bco\ at three different temperatures are shown along with Rietveld refinement results at (a) 30~K ($T > T_{N3}$), (b) 10~K ($T_{N1} < T < T_{N2,N3}$), and (c) 2~K ($T < T_{N1}$). Best fits to the experimental data was obtained for the superspace group $P21nm.1^{\prime}(00g)0s0s$ at 10~K and $P21/b.1^{\prime}[b](00g)00s$ at 2~K.
		The insets show a section of diffraction pattern where few additional Bragg peaks are observed compared to the 30 K data. Arrangement of Cr$^{3+}$ spins viewed along $b$ direction at (d) 10~K and (e) 2~K.}
\end{figure*}
\indent 
In figure~\ref{fig-npd2}(a--c) we present the neutron powder diffraction patterns of \bco\ collected at three exemplary temperatures, 30~K (paramagnetic), 10~K ($T_{N1} < T < T_{N2,N3}$), and 2~K ($T < T_{N1}$), along with the Rietveld refinement results of nuclear and magnetic structures.
At 30~K, which is slightly above $T_{N3}$, the diffraction pattern shows purely nuclear contributions as can be seen from Figure~\ref{fig-npd2} (a). 
At 10~K and 2~K, the strong magnetic Bragg peak at $d$ = 10~{\AA} is a conspicuous feature in the diffractograms, see Figure~\ref{fig-npd2} (b), (c). 
The low-temperature diffraction data below $T_{N2,N3}$ was refined using the magnetic superspace group approach \cite{perez2015symmetry}.
\begin{table}[!t]
	\centering
	\caption{Structural data of magnetic unit cell of BeCr$_2$O$_4$ at 10~K and 2~K, obtained from the Rietveld refinement of neutron diffraction data shown in Figure~\ref{fig-npd2}.} \vspace{5pt}
	\label{tab-npd-lowT}
	\begin{tabular*}{\columnwidth}{@{\extracolsep{\fill}}cccccc} \hline \hline
		\multicolumn{2}{l}{Temperature (K)} &\multicolumn{2}{c}{$10$} & \multicolumn{2}{c}{$2$} \\ \hline
		\multicolumn{2}{l}{Modulation vector ($\textbf{k}$)} & \multicolumn{2}{c}{$(0,0,0.090(1))$} & \multicolumn{2}{c}{$(0,0,0.908(1))$} \\ \hline
		\multicolumn{2}{l}{Superspace group} &\multicolumn{2}{c}{$P21nm.1^{\prime}(00g)0s0s$} & \multicolumn{2}{c}{$P21/a.1^{\prime}[a](00g)00s$} \\ \hline
		\multicolumn{2}{l} {Lattice parameters} & & & &   \\
		\multicolumn{2}{l}{$a$~({\AA})} & \multicolumn{2}{c}{$4.6812(3)$} & \multicolumn{2}{c}{$9.3320(7)$} \\
		\multicolumn{2}{l}{$b$~({\AA})} & \multicolumn{2}{c}{$10.0581(1)$} & \multicolumn{2}{c}{$11.0577(8)$} \\
		\multicolumn{2}{l}{$c$~({\AA})} & \multicolumn{2}{c}{$5.8076(1)$} & \multicolumn{2}{c}{$5.7885(4)$} \\
		\multicolumn{2}{l}{$V$~({\AA$^3$})} & \multicolumn{2}{c}{$273.45(1)$} & \multicolumn{2}{c}{$541.5(1)$} \\
		\multicolumn{2}{l}{$\gamma~(^\circ)$} & \multicolumn{2}{c}{$90$} & \multicolumn{2}{c}{$114.956(2)$}\\ \hline
		\multicolumn{2}{l} {Atomic positions} & & & \\
		& $x/a$ & \multicolumn{2}{c}{$0$} & \multicolumn{2}{c}{$0$}  \\
		\multicolumn{1}{l}{Cr1} & $y/b$ & \multicolumn{2}{c}{$0$} & \multicolumn{2}{c}{$0$}  \\
		& $z/c$ & \multicolumn{2}{c}{$0$} & \multicolumn{2}{c}{$0$}  \\ \cline{2-6}
		& $x/a$ & \multicolumn{2}{c}{$0.508(1)$} & \multicolumn{2}{c}{$-0.5$}  \\
		\multicolumn{1}{l}{Cr2} & $y/b$ & \multicolumn{2}{c}{$0.224(1)$} & \multicolumn{2}{c}{$-0.5$}  \\
		& $z/c$ & \multicolumn{2}{c}{$0.75$} & \multicolumn{2}{c}{$0$}  \\ \cline{2-6}
		& $x/a$ & \multicolumn{2}{c}{$-0.489(1)$} & \multicolumn{2}{c}{$-0.123(1)$}  \\
		\multicolumn{1}{l}{Cr3} & $y/b$ & \multicolumn{2}{c}{$-0.235(1)$} & \multicolumn{2}{c}{$-0.771(1)$}  \\
		& $z/c$ & \multicolumn{2}{c}{$-0.75$} & \multicolumn{2}{c}{$0.75$}  \\ \cline{2-6}
		& $x/a$ & \multicolumn{2}{c}{$-$} & \multicolumn{2}{c}{$-0.620(1)$}  \\
		\multicolumn{1}{l}{Cr4} & $y/b$ & \multicolumn{2}{c}{$-$} & \multicolumn{2}{c}{$-0.267(1)$}  \\
		& $z/c$ & \multicolumn{2}{c}{$-$} & \multicolumn{2}{c}{$0.75$}  \\ \hline
		\multicolumn{2}{l} {$R$--factors} & & &   \\
		& $R_\mathrm{obs}$ & \multicolumn{2}{c}{$3.84$} & \multicolumn{2}{c}{$4.64$}  \\
		& $wR_\mathrm{obs}$ & \multicolumn{2}{c}{$5.08$} & \multicolumn{2}{c}{$6.88$}  \\ \hline \hline
	\end{tabular*}
\end{table}
Starting with the parent crystal structure ($Pbnm$) the magnetic modulation is found to be incommensurate with a modulation vector $\textbf{k}_1$~=~(0, 0, 0.090(1)) in the temperature rage $T_{N1} < T_{N2,N3}$. 
At 2~K, in addition to the aforementioned reflections, a new set of reflections appear that can be indexed with a modulation vector (0.5, 0.5, 0.0) with the parent setting. 
The magnetic structure below $T_{N1}$ was refined using superspace group approach by using a larger monoclinic supercell (Table~\ref{tab-npd-lowT}). 
With this approach all the magnetic reflections could be indexed with a single modulation vector $\textbf{k}_2$~=~(0, 0, 0.908(1)).
The best fits to the experimental data were obtained with the magnetic superspace groups $P21nm.1^{\prime}(00g)0s0s$ at 10~K and $ P21/a.1^{\prime}[a](00g)00s$ at 2~K. 
The modulated unit cell parameters are given in Table~\ref{tab-npd-lowT}. 
The resulting magnetic supercell is pictorially represented in Figure~\ref{fig-npd2} (d) and (e) as projections along $ac$ plane at 10~K and 2~K respectively. 
The Cr$^{3+}$ spins form a cycloid with the periodicity extending to about 65~{\AA}.
%
%
\section{Discussion}
Several types of magnetic ordering can lead to the development of ferroelectric polarization in type-II multiferroics \cite{tokura2010multiferroics, tokura2014multiferroics}. 
Cycloidal, conical, screw, collinear antiferromagnetic and weak ferromagnetic order are all known to result in non-zero electric polarization. 
TbMnO$_3$, DyMnO$_3$, MnWO$_4$, CuO, \cite{kimura2003magnetic, goto2004ferroelectricity, taniguchi2006ferroelectric, kimura2008cupric} CoCr$_2$O$_4$, ZnCr$_2$Se$_4$, \cite{yamasaki2006magnetic, siratori1980method} CuFeO$_2$, CuCrO$_2$, \cite{kimura2006inversion, seki2008spin} DyMn$_2$O$_5$, \cite{hur2004electric} DyFeO$_3$, GdFeO$_3$ \cite{tokunaga2008magnetic, tokunaga2009composite} are examples of compounds that display the different types of magnetic order mentioned above and consequently develop electric polarization, either spontaneously or with the application of an external magnetic field. 
When the spins in a crystal lattice form a cycloidal modulation along a crystallographic direction, a macroscopic polarization, $\textbf{\emph{P}}~=~a\sum \textbf{\emph{e}}_{ij} \times \left(\textbf{\emph{S}}_i \times \textbf{\emph{S}}_j\right)$ is generated; where $a$ is determined by spin-orbit coupling, exchange and spin-lattice interactions, $e_{ij}$ is the unit vector connecting the neighbouring spins $S_i$ and $S_j$.
This type of polarization can be explained using the spin-current model or the Katsura--Nagaosa--Balatsky model \cite{katsura2005spin}
which is derived from an inverse Dzyaloshinskii-Moriya model \cite{sergienko2006role}. 
\\
\indent 
Perovskite manganites were the first to be studied from the perspective of cycloidal spin order. 
For example, in TbMnO$_3$ the Mn$^{3+}$ spins undergo a collinear sinusoidal antiferromagnetic order below $T_N \approx$ 41~K with a wave vector $\bf k$ = (0, $k_s$, 1) in $Pbnm$ space group \cite{kimura2003magnetic}. 
The value of incommensurate wave vector $k_s$ at $T_N$ was found to be 0.295 which gradually evolves with decreasing temperature, reaching the value of 0.28 at 30~K and remains constant below this temperature. 
\bco\ orders at a temperature similar to that of TbMnO$_3$ in which the Mn$^{3+}$ spins locks-in at the commensurate value of $k_s$. 
In \bco\ we find that the modulation vector is (0,~0,~0.090(1)) with the $Pbnm$ space group in the temperature range 26 -- 7.5~K and changes to (0,~0,~0.908(1)) below 7.5~K with a larger monoclinic supercell. 
Albeit some similarities, TbMnO$_3$ and \bco\ have different magnetic lattices; the former has a three-dimensional magnetic lattice while the latter possesses a quasi-two-dimensional sawtooth structure. 
Sawtooth multiferroics are very rare in the literature and thus only seldom investigated. An exception is the well studied compound Mn$_2$GeO$_4$, which also has an olivine-based crystal structure \cite{honda2012structure}. 
This compound exhibits successive magnetic phase transitions at $T_{N1}$~=~47~K, $T_{N2}$~=~17~K and $T_{N3}$~=~5.5~K similar to \bco\ which also shows three anomalies below 30~K.
A substantial amount of lattice distortion was found at the transition temperatures in Mn$_2$GeO$_4$ where Mn$^{2+}$ spins form sawtooths along the $b$ axis. 
Different from the cycloidal order in the classic example of TbMnO$_3$ or in \bco, a commensurate magnetic order with modulation vector $\bf k$ = (0,~0,~0) is established in Mn$_2$GeO$_4$ at high temperatures, which then transforms to a multi-component magnetic structure when combined with an incommensurate ordering vector, (0.136(2),~0.211(2),~0) at low temperature. 
The spontaneous polarization developed in these compounds is ascribed to an incommensurate spiral order. 
\bco, although a sawtooth lattice-like Mn$_2$GeO$_4$, is found in the present work to adopt an incommensurate magnetic order below the magnetic phase transition temperature at $\sim$26~K. 
The propagation vector of $\bf k$ = (0, 0, 0.090(1)) remained a constant down to $T_{N1}$. The arrangement of Cr$^{3+}$ spins at the lowest temperature of our study is shown in Figure~\ref{fig-npd2}~(d).
\\
\indent
Multiple nearby magnetic transitions are similarly observed in another sawtooth lattice multiferroic, Cu$_3$Nb$_2$O$_8$ \cite{johnson2011cu}.
Very similar to \bco, Cu$_3$Nb$_2$O$_8$ also shows a double-peak in the specific heat \cite{johnson2011cu}.
This compound crystallizes in centrosymmetric triclinic symmetry, $P\overline{1}$, but orders magnetically at 26~K and develops electric polarization stemming from the cycloidal magnetic order at 24 K.
The magnetic order in the polar phase of Cu$_3$Nb$_2$O$_8$ has a propagation vector of $\bf k$ = (0.4879, 0.2813, 0.2029).
An electric polarization of magnitude, 17.8~$\mu$Cm$^{-2}$ is observed almost perpendicular to the plane of rotation of the spins.
This is in contradiction with the model predictions for electrical polarization in cycloidal multiferroics that should be contained within the plane in case of coplanar spins.
In the high symmetry general cases, where the polarization is developed as a result of the DM interaction, the polarization develops in a direction perpendicular to both the magnetic propagation vector and the normal to the plane of rotation of the spins \cite{mostovoy2006ferroelectricity}.
Contrary to the predictions for the cycloidal magnets, the polarization in Cu$_3$Nb$_2$O$_8$ lies perpendicular to the plane of the magnetic moments.
This is explained as due to the coupling between a macroscopic structural rotation and the magnetically induced structural chirality \cite{johnson2011cu}.
These examples imply that a closer and more detailed understanding of the structural distortions and rotations in sawtooth magnets, like \bco, is extremely instructive.
\\
\indent
The multiple phase transitions observed in \bco\ at low temperature leads to a slightly more complex magnetic field-temperature phase diagram compared to that of atacamite Cu$_2$Cl(OH)$_3$, which also exhibits a sawtooth arrangement of spins \cite{heinze2021magnetization}. 
The H-T phase diagram of this atacamite was determined up to high magnetic fields, up to 600~kOe. 
A complex magnetic phase evolution is seen in single crystal samples of atacamite, where at 315~kOe, the magnetization isotherms attain a plateau.
Interestingly, this plateau is argued not to be related to the magnetization plateau predicted in the magnetization of a spin-half sawtooth lattice, that is the $\Delta$ chain \cite{kubo1993excited}. 
Below the antiferromagnetic anomaly observed in the atacamite at 8.5~K, the compound enters a magnetically ordered state with a propagation vector equal to (1/2, 0, 1/2). 
It can be pointed out here that a weak low--temperature anomaly is observed in the specific heat of atacamite at approximately 5~K. 
This indicates a complex low-temperature magnetic phase, similar to what we are reporting here in the case of \bco. 
Moreover, a small value of magnetic entropy equal to 0.65Rln(2) is observed in atacamite at the transition temperature, hinting at the importance of the role of magnetic frustration. 
Judging from the magnetization curves presented in the current work on polycrystalline samples, it is clear that \bco\ displays magnetic field-dependent phase transitions below $T_{N3}$. 
Experiments on single crystals are essential to reveal anisotropy effects in the magnetic response as well as the type of magnetic ordering in the temperature range $T_{N3} > T > T_{N2}$. 
High-magnetic field response, in the range of 600~kOe or more, will be beneficial to examine if a magnetization plateau is also present in \bco\ and, if so, how it compare with atacamite. 
Thus it is our immediate plan to investigate the high-field magnetic response of the magnetic structures of this compound using neutron diffraction experiments on large single crystals. 
Definitely, a comprehensive characterization of the electrical polarization and magnetoelectric effect in \bco\ is highly warranted to explore multiferroicity in sawtooth lattices.
\\
\indent
The bond angle Cr--O--Cr determined for \bco\ is conducive for strong superexchange mechanism between the Cr moments.
Strong superexchange is known to be influential
in bringing about induced-multiferroicity 
in compounds like CuO and BiFeO$_3$ \cite{kimura2008cupric, shen2021non}.
The systematic correlation between the
Cu--O--Cu bond angle and the magnetic
exchange interaction $J$ is understood
in the cuprates \cite{shimizu2003magnetic}.
The importance of superexchange
interaction in multiferroic compounds
is highlighted in recent studies on the
popular multiferroic compound, BiFeO$_3$
\cite{shen2021non}.
The stereochemical activity of Bi's
lone pair electrons alone cannot explain
the observed electric polarization of
BiFeO$_3$.
It is proposed that the Fe 3$d$ orbitals
render the compound a charge-transfer
insulator and the Fe--O--Fe superexchange
enhances the developed electric polarization.
The validity of this mechanism in the present
compound \bco\ and the extent to which it
enhances the electric polarization remains to be
investigated in detail.

\section{Conclusions}
Our study reveals multiple magnetic phase transitions in \bco\ present below 26~K, which is identified as the magnetic phase transition temperature of this compound.
A double-peak feature is observed (25~K and 26~K) in the physical properties of \bco.
The magnetic field--temperature phase diagram indicates two antiferromagnetic phases, with possibly a complex phase region between the double-peak transitions.
A noncollinear magnetic structure is estimated at low temperature for \bco\ which supports the development of electric polarization through breaking of the inversion symmetry upon establishment of magnetic order.
A re-investigation of low-temperature magnetic phases, electric polarization, and high-magnetic field responses of \bco\ will be highly rewarding in the future.
%
%
\section{Acknowledgements}
Use of the Advanced Photon Source at Argonne National Laboratory was supported by the U. S. Department of Energy, Office of Science, Office of Basic Energy Sciences, under Contract No. DE-AC02-06CH11357. 
A portion of this research used resources at the  Spallation Neutron Source, a DOE Office of Science User Facility operated by the Oak Ridge National Laboratory.
This work has been supported in part by the Croatian Science Foundation under Project No. IP-2020-02-9666. 
CMNK acknowledges the support of project Cryogenic Centre at the Institute of Physics - KaCIF, co-financed by the Croatian Government and the European Union through the European Regional Development Fund - Competitiveness and Cohesion Operational Programme (Grant No. KK.01.1.1.02.0012).
The work at the TU Wien was supported by the European Research Council (ERC Consolidator Grant No. 725521).
The work at AGH University of Science and Technology was supported by the National Science Centre, Poland, grant OPUS: UMO-2021/41/B/ST3/03454, the Polish National Agency for Academic Exchange under 'Polish Returns 2019' Programme: PPN/PPO/2019/1/00014, and the subsidy of the Ministry of Science and Higher Education of Poland.
%

%
\end{document}